# Product Line Annotations with UML-F

Wolfgang Pree[1], Marcus Fontoura[2], and Bernhard Rumpe[3]

[1] Department of Computer Sciences (guest), Univ. of California, Berkeley,
pree@eecs.berkeley.edu
[2] IBM Almaden Research Center, 650 Harry Road,
95120 San Jose, CA, U.S.A
fontoura@almaden.ibm.com
[3] Software and Systems Engineering, Munich University of Technology,
D-80290 Munich, Germany,
rumpe@acm.org
Internet: http://uml-f.net/

**Abstract.** The Unified Modeling Language (UML) community has started to define so-called profiles in order to better suit the needs of specific domains or settings. Product lines[1] represent a special breed of systems—they are extensible semi-finished pieces of software. Completing the semi-finished software leads to various software pieces, typically specific applications, which share the same core. Though product lines have been developed for a wide range of domains, they apply common construction principles. The intention of the UML-F profile (for framework architectures) is the definition of a UML subset, enriched with a few UML-compliant extensions, which allows the annotation of such artifacts. This paper presents aspects of the profile with a focus on patterns and exemplifies the profile's usage.

## 1   What Is a UML Profile?

The UML is a large and regrettably complex language. Still, there are many requests to explicitly represent additional features that cannot be described comfortably using the current version of the UML. Therefore, the UML provides mechanisms, in particular *stereotypes* and *tagged values*, which allow extensions. These extensions may be defined and grouped in so-called *profiles*.

Thus, a UML profile is defined as an extension of the standard UML with specific elements. A profile provides new notational elements and specializes the semantics of some elements. It may also restrict the use of UML elements. For example, [2] describes in further detail the profiling mechanism and a useful extension of it, called *prefaces*.

A UML profile may target a specific application domain. UML-RT, the real-time profile, is one prominent example. Other profiles may provide tool-specific exten-sions. For example, these might shape the UML so that it is better suited for modeling Web-based systems, as the one described in [1]. A Java™ profile would restrict the UML to single-class inheritance. The UML-F profile, which is described in detail in [4], supports product line annotations.

---

[1] We use the terms product line and framework synonymously.

## 2   A Selection of Basic UML-F Tags

Though the UML version 1.3 already lists 47 stereotypes and 7 tagged values [6] and version 1.4 increased these numbers considerably, only a small number of them are particularly useful for product line annotations. This section picks out some of the tags[2] introduced by the UML-F profile for that purpose.

### 2.1   Product Line and Application Classes

Many product lines come together with prefabricated application classes that do not belong to the product line itself. These additional classes can be studied in order to understand the standard usage of a framework by examining and adapting their code, whereas the product line classes themselves are usually not subject to change.

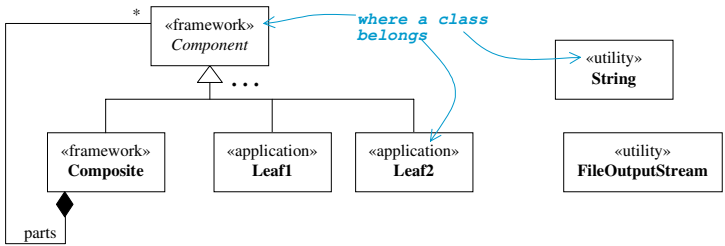

**Fig. 1.** «framework», «application» and «utility» Tags

The «application» tag marks application-specific classes. During the product line adaptation process, this tag may mark newly introduced classes as well. The «framework» tag marks classes and interfaces belonging to the product line. A third category of classes belongs to the utility level, as these classes are provided as basic classes by utility libraries or the runtime system. These may be tagged by «utility». Figure 1 exemplifies their usage. If a package is marked with one of these tags, then all of its classes and interfaces are implicitly marked as such. Table 1 summarizes the meaning of these UML-F tags.

---

™  Java and all Java-based marks are trademarks of Sun Microsystems, INC. in the U.S. and other countries.
[2]  UML-F provides a slightly simplified mechanism that unifies standard UML stereotypes and tagged values to UML-F *tags* [4].

**Table 1.** UML-F Class Tags for Discerning Between Framework and Application Components

| Tag-name | Applies to | Value type | Description |
|---|---|---|---|
| «application» | Class, Package Interface | Bool | The class/interface/package does not belong to a framework, but to an application. |
| «framework» | Class, Package Interface | Bool | The class/interface/package belongs to the framework. |
| «utility» | Class, Package Interface | Bool | The class/interface/package belongs to a utility library or the runtime system. |

## 2.2   Completeness and Abstraction

The following UML-F tags deal only with the visual representation of elements; they do not define any properties of the annotated modeling elements.

The standard UML provides the ellipsis (…) to mark the omissions of attributes and methods. However, we also find it useful to be able to mark elements as complete. Therefore, we propose the UML-F © tag to mark completeness and the ellipsis tag to explicitly mark incompleteness. In accordance with standard UML, the ellipsis tag is the default. This means that class diagrams and all their elements are considered incomplete unless explicitly marked as complete via the use of the © tag.

Figure 2 shows three representations of the class Human. The first is complete, whereas the second partly omits the attribute compartment, and the third omits both the attribute and method compartments.

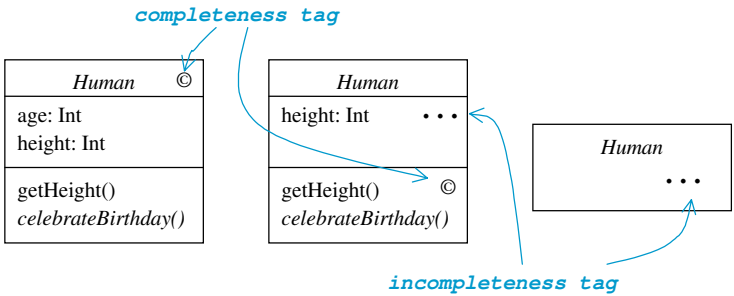

**Fig. 2.** Three Representations of the Same Class

## 3    UML-F Pattern Tags

Many patterns written up in the pioneering pattern catalog [5] by the *Gang-of-Four* (GoF—the four authors Gamma, Helm, Johnson and Vlissides) are product lines that rely on a few framework construction principles. This section discusses the relationship between the construction principles and the GoF catalog patterns, and introduces the UML-F tags for annotating both the framework construction principles as well as design patterns. Instead of listing the UML-F tags for each of the construction principles and patterns, we explain and exemplify how to derive the UML-F tags in a straightforward manner from their static structure.

### 3.1    Unification Principle – Adaptation by Inheritance

Hook methods can be viewed as placeholders that are invoked by more complex methods that are usually termed *template methods*[3] [3, 5, 7, 8]. The simple idea behind hook methods is that overriding hooks allows changes of the corresponding template method's behavior without having to touch the source code of the class to which the template method belongs.

The essential set of framework construction principles can be derived from considering all possible combinations between template and hook methods within one class or in two classes. The reason why this becomes feasible is that the abstract concepts of templates and hooks fade out domain-specific semantics to show the clear means of achieving extensibility in object-oriented technology. The UML-F tags for explicitly marking templates and hooks are «template» and «hook». In the sample UML-F diagram in Figure 3, the `convert` () method in the class CurrencyConverter is a template method invoking `round` () as its hook.

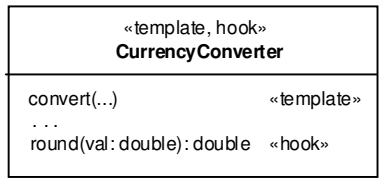

**Fig. 3.** UML-F Tags Annotating Template and Hook Methods

These UML-F tags can be used not only to mark methods, but also to mark classes and interfaces. Attaching the «hook» tag to a class or interface means that it contains a hook method. The «template» tag has the analogous meaning, though it only makes sense to attach it to classes and not to interfaces, since interfaces cannot provide method implementations. Figure 3 attaches both tags to the CurrencyConverter class, applying the UML-F tag rule that allows the listing of multiple tags.

---

[3]  Template methods must not be confused with the C++ template construct, which has a completely different meaning.

The template-hook combination in which the template method and its corresponding hook methods reside in the same class, (as in the case of Figure 3) corresponds to the Unification construction principle. The general description of that principle calls the class TH (see Figure 4). The recipe for deriving the tags is to concatenate the construction principle name with each of the elements of the general structure of the construction principle. Thus, three UML-F tags annotate the Unification construction principle in a framework:

- «Unification–TH» marks the class
- «Unification–t» marks the template method
- «Unification–h» marks the hook method(s)

As a short cut, we suggest using the tags «Unif–TH», «Unif–t» and «Unif–h». Compared to the bare-bones template and hook tags, the explicit stating that the Unification construction principle underlies a certain aspect of the framework provides more semantic information. In particular, one who is familiar with the Unification construction principle might, for example, infer the degree of flexibility associated with that construction principle: Adaptations have to be accomplished in subclasses and thus require an application restart. This illustrates the layered structure of the UML-F tags – more "semantic-rich" tags can be defined in terms of more basic ones. This layered structure of the UML-F profile is further described in section 3.2.

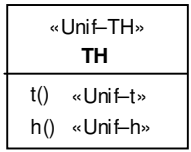

**Fig. 4.** Static Structure of the Unification Construction Principle

Figure 5 applies the Unification tags to annotate the CurrencyConverter class. In UML-F, any name can be defined for a group of tags. We chose the name Rounding for the Unification construction principle in that case. Although Figure 5 and Figure 3 represent the same aspect of the product line, Figure 5 makes the underlying construction principle explicit, while Figure 3 uses the more basic template-hook tags.

Conceptually, UML-F tags that correspond to the static structure of a framework construction principle provide a means for pinpointing the methods and classes in a product line that apply a particular design. Figure 6 illustrates that aspect. The arrows express the mapping of the structural components of the Unification construction principle to their manifestation in a certain part of a framework.

The Separation construction principle derives from the Unification construction principle by moving the hook method to a separate class H. The class T containing the template method has an association to H. The template method in T invokes the hook method in H through this association. The difference to the Unification construction principle is that the behavior of T can be changed at runtime by plugging in a specific H instance. The UML-F annotation is analogous to the Unification principle and discussed in detail in [4].

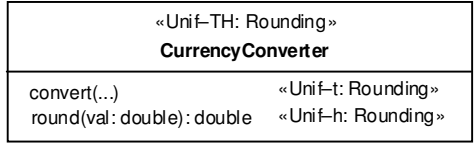

**Fig. 5.** UML-F Annotation of a Sample Application of the Unification Construction Principle

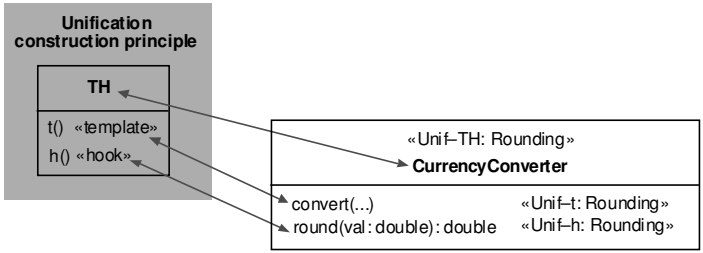

**Fig. 6.** Rationale Behind the Unification UML-F tags

### 3.2   UML-F Tags for Design Patterns

Analogous to the Unification and Separation construction principles, the structure of a design pattern determines the particular set of UML-F tags. In the case of the GoF patterns, each pattern description has a section labeled Structure that shows a class diagram. The class and method names in such a diagram, together with the pattern name, form the set of UML-F tags: «PatternName–methodName», «PatternName–ClassName» and for potential future Java- or C#-based versions of the GoF catalog or other pattern catalogs: «PatternName–InterfaceName». In occasions where associations and attributes play a role, tags of the form «PatternName-associationLabel» and «PatternName-attributeName» are present as well. This section illustrates that description scheme for the GoF pattern Factory Method.

Consider the layered relationship between the UML-F tags sets[4] for the design patterns, construction principles, and the template and hook tags (see Figure 7). The essential framework construction principles Unification, Separation, Composite, Decorator and Chain-Of-Responsibility represent the possible combinations of templates and hooks in one or two classes/interfaces. The patterns Composite, Decorator and Chain-Of-Responsibility are identical to those core framework construction principles that result from combinations of templates and hooks via inheritance. We suggest the names and structure of these three patterns as the basis from which to derive the UML-F tag sets.

---

[4] The tags of a construction principle or pattern form one tag set. For example, the three tags of the Unification principle form the Unification tag set (<<Unif-TH>>, <<Unif-t>>, and <<Unif-h>>).

Though the other GoF framework patterns and domain-specific patterns rely on either the Unification or Separation of templates and hooks, product line developers who need to express the richer semantic information inherent in these patterns, annotate a product line by means of the corresponding UML-F tag sets. The application of one of these patterns expresses the intended use and adaptation possibilities beyond the static structure to which a pattern is often reduced. The semantics of a pattern tag are defined through its structural and behavioral constraints, as well as its intended use.

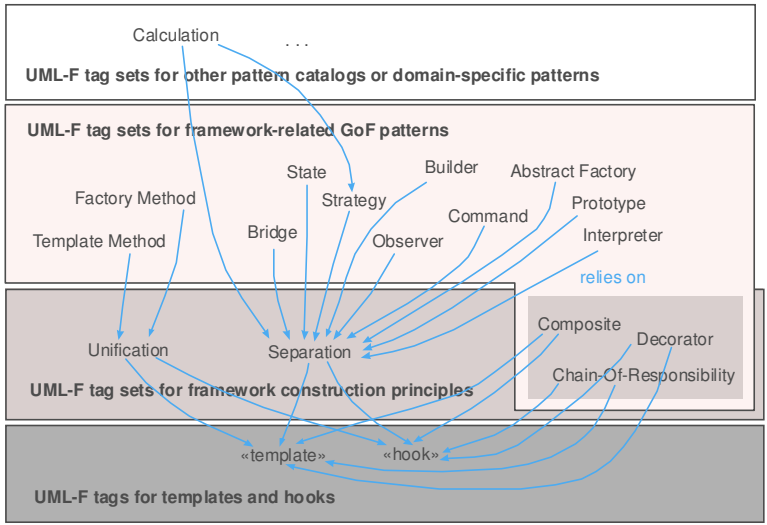

**Fig. 7.** Layers of UML-F Tag Sets

### 3.2.1   UML-F Tags for the Factory Method Pattern.

Figure 8 shows the static structure of the Factory Method pattern according to the GoF catalog [5]. Note that the diagrams in the GoF catalog adhere to the Object Modeling Technique (OMT) [8] notation which significantly influenced the UML, but which is no longer used in its original form. All the diagrams in this paper adhere to the UML notation.

According to the static structure of the Factory Method pattern[5], the UML-F tag set consists of these tags[6]:

- «FacM–Creator»
- «FacM–facM»
- «FacM–anOp»
- «FacM–Product»
- «FacM–ConcreteProduct»

---

[5] The tags adopt the Java convention of using uppercase first letters in class/interface names and lowercase first letters in method names. The use of the *italics* style indicates an abstract class, an interface or an abstract method.

[6] We suggest the abbreviations FacM for Factory Method and anOp for anOperation.

- «FacM– ConcreteCreator»
- «FacM–facM»

One could argue that the subclasses of Creator and Product are in some cases not relevant to documenting that pattern. So an alternative, shorter list of UML-F tags would be: «FacM–Creator», «FacM–*facM*», «FacM–anOp», «FacM–Product».

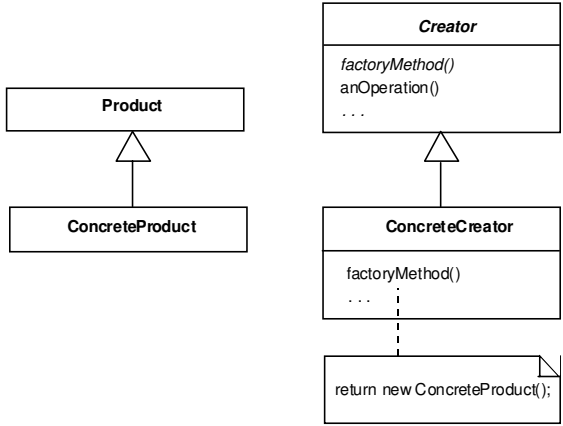

**Fig. 8.** Structure of the Factory Method Pattern (adapted from [5])

Figure 9 attaches the UML-F template and hook tags to the methods of the Creator class. This illustrates in detail the relationship between the tag sets of the Factory Method Pattern and the Unification construction principle. It also demonstrates the additional semantic information provided by the Factory Method pattern. For example, the Unification construction principle does not deal with a Product class.

### 3.3   Hooks as Name Designators of Pattern Catalog Entries

Hook methods form the points of predefined refinement that we call variation points or hot spots [7]. Product line adaptation takes place at these variation points. Depending on the hook placement and template-hook method combination used, more or less flexibility can be achieved.

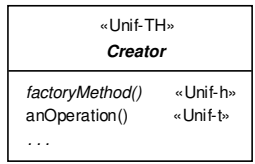

**Fig. 9.** Application of the Unification Construction Principle in the Factory Method Pattern

Every product line incorporates the two essential construction principles Unification and Separation, no matter how simple or how complex the particular template and hook methods are. Fine-grained classes apply the same construction principles as complex classes in order to introduce flexibility. They differ only in the granularity, the hook's semantics (often expressed in a hook's name), the number of defaults provided for a hook, and the number of template-hook pairs. Thus, we can take a fresh look at the design pattern catalog, that is, the 23 patterns published in [5].

Many entries in the pattern catalog can be regarded as small product lines[7], consisting of a few classes, which apply the essential construction principles in various more or less domain-independent situations. These catalog entries are helpful when designing product lines, and they illustrate typical hook semantics. In general, the names of the catalog entries are closely related to the semantic aspects that are kept flexible by the provided hooks.

A significant portion of the framework-centered pattern catalog entries relies on a separation of template and hooks, that is, on the basic Separation principle. The catalog pattern Bridge discusses the abstract-coupling mechanism generically. Other catalog entries that use template-hook separation introduce more specific semantics for their hooks: Abstract Factory, Builder, Command, Interpreter, Observer, Prototype, State and Strategy. The names of these catalog patterns correspond to the semantics of a particular hook method or the corresponding class.

### 3.4     UML-F Tags for Domain-Specific Patterns

Product lines contain numerous patterns that are not general enough to be published in pattern catalogs, but that rely on one of the essential framework construction principles. In many situations, it might be useful to introduce UML-F tag sets that explicitly refer to these domain-specific patterns. The definitions of these domain-specific UML-F tag sets work in the same way as for the pattern tags. The structure of a domain-specific pattern defines the tags. The structure of a domain-specific pattern should also be annotated, either by GoF pattern tags or the tags of the core construction principles. This ensures an explanation of the domain-specific pattern in terms of already understood designs.

## 4     Conclusions

The intention of the UML-F profile is the identification of a UML subset, enriched with UML-compliant extensions, which allows the annotation of product lines. Overall, the presented selection of key aspects of the UML-F profile pursues the following goals:

---

[7] These include Template Method, Factory Method; Bridge, Abstract Factory, Builder, Command, Interpreter, Observer, Prototype, State, Strategy; and Composite, Decorator, Chain-of-Responsibility.

1. UML-F provides the notational elements to precisely annotate and document well-known design patterns. Only a rather limited UML support currently exists for that purpose.
2. UML-F is itself in the spirit of frameworks—straightforward extensibility is the key to providing a suitable means for documenting any framework pattern including the future ones.
3. UML-F comprises a lean, mnemonic set of notational elements.
4. UML-F relies on the UML standard, that is, the extensions should be defined on the basis of the existing UML extension mechanisms.
5. The notational elements are adequate for being integrated in UML tool environments. For example, tools should be able to create hyperlinks between annotated framework patterns and the corresponding online pattern documentation.

More profiles will be standardized by the OMG in the future; sound proposals from various communities will get the process of defining and standardizing UML profiles started. In that sense, UML-F sets the stage for the UML profile for product lines.